\documentclass[reprint,amsmath,amssymb,aps]{revtex4-2}

\usepackage{tabularx}
\usepackage{multirow}
\usepackage{graphicx}
\usepackage{dcolumn}
\usepackage{bm}

\usepackage{xcolor}

\usepackage{tikz}
\usetikzlibrary{arrows.meta}

\definecolor{plum}{rgb}{0.36078, 0.20784, 0.4}
\definecolor{chameleon}{rgb}{0.30588, 0.60392, 0.023529}
\definecolor{cornflower}{rgb}{0.12549, 0.29020, 0.52941}
\definecolor{scarlet}{rgb}{0.8, 0, 0}
\definecolor{brick}{rgb}{0.64314, 0, 0}
\definecolor{sunrise}{rgb}{0.80784, 0.36078, 0}
\definecolor{lightblue}{rgb}{0.15,0.35,0.75}
\definecolor{carolina}{RGB}{153, 186, 221}
\tikzstyle{axisarrow} = [-{Latex[inset=0pt,length=5pt]}]

\begin{document}

\preprint{APS/123-QED}

\title{On the Hubble expansion in a Big Bang quantum cosmology}

\author{Maurice H.P.M. vanPutten,$^{1,2}$}
\email{mvp@sejong.ac.kr}
\affiliation{$^1$Physics and Astronomy, Sejong University, 209 Neungdong-ro, Seoul, South Korea,} 
\affiliation{$^2$INAF-OAS Bologna via P. Gobetti 101 I-40129 Bologna Italy,
Italy} 

\date{\today}

\begin{abstract}
The Hubble expansion of the Universe is considered in the classical limit of a Big Bang quantum cosmology. In an IR-consistent coupling to the the bare cosmological constant, we infer a dark energy as a relic of the Big Bang by loss of time-translation invariance on a Hubble time-scale.
This dark energy is identified with the trace $J$ of the Schouten tensor permitting an analytic solution $H(z)$. Anchored by the {\em Baryonic Accoustic Oscillations}, 
$J$CDM predicts a Hubble constant $H_0=\sqrt{6/5}\,H_0^\Lambda$ alleviating $H_0$-tension between the Local Distance Ladder and $H_0^\Lambda$ in $\Lambda$CDM, whose dark energy $\Lambda$ is a constant. 
{Emulated by $w(a)\Lambda$CDM}, 
a CAMB analysis shows a $J$CDM fit to the {\em Planck} 2018 $C_l^{TT}$ power spectrum on par with $\Lambda$CDM with small positive curvature consistent with {\em Planck}-$\Lambda$CDM with no extra relativistic degrees of freedom. In late-time cosmology, $J$CDM is also consistent with the BAO recently measured by DESI. { $J$CDM offers a novel framework to address $H_0$-tension, predicting background quantities consistent with the uncertainties in BAO measurements and early-Universe observations.} It predicts a deceleration parameter $q_0\simeq-1$, that may be tested with upcoming low-redshift galaxy surveys.
\end{abstract}


\maketitle

{\em Introduction.} From dawn at the Big Bang to the present epoch, the Universe has been expanding for a Hubble time with the formation of the large-scale structure in galaxies on smaller scales. This process evolves in weak gravitation largely at the cosmological de Sitter scale of acceleration $a_{dS}=cH$, where $c$ is the velocity of light and $H$ is the Hubble rate of expansion.
By the Copernicus principle, the Universe on the largest scales is  believed to be homogeneous and isotropic, described in the Friedmann-Lema\^itre-Robertson-Walker (FLRW) line-element \cite{san61}
\begin{eqnarray}
    ds^2 = a^2\eta_{ab}dx^adx^b
    \label{EQN_F}
\end{eqnarray}
by conformal scaling $a$ of the Minkowski metric $\eta_{ab}$.
Equivalently,  
$ds^2=-c^2dt^2+a(t)^2\left(dx^2+dy^2+dz^2\right)$ expresses $H=\dot{a}/a$ as a function of cosmic time $t$ with Hubble radius $R_H=c/H$ and deceleration parameter $q=-a\ddot{a}/\dot{a}^2$. The latter equals $q(z)=-1 + (1+z)H^{-1}H^\prime$ in terms of redshift $z$, where $a/a_0=1/(1+z)$ normalized to $a_0$ today \cite{san61}. 

Crucially, the Hubble parameter $H_0=H(0)$ points to a Big Bang, a time $H_0^{-1}$ in the past \cite{agh20}. 
While a theory of quantum cosmology remains elusive \citep{agr18}, this Big Bang breaks time-translation invariance that, on a Hubble time, introduces a small energy scale $\epsilon \sim H\hbar$ per degree of freedom \citep{van20}. 
The dimension of phase space within a radius $r\le R_H$ is finite by the Bekenstein bound $N=A_{p}/4$ by the area $A_p=4\pi r^2/l_p^2$ in Planck units, $l_p=\sqrt{G\hbar/c^3}$ \cite{bek81}, given Newton's constant $G$ and Planck's constant $\hbar$. Including a factor of $1/2\pi$, preserving consistency with the first law of thermodynamics \citep{pad05,van24e}, $\epsilon=H\hbar/2\pi$
\citep{gib77} is an energy scale of the vacuum of de Sitter space, pointing to a relic heat of the Big Bang
\begin{eqnarray} 
 Q\simeq N\epsilon
\label{EQN_Q}
\end{eqnarray}
distinct from the semi-classical UV-catastrophe \citep{wei89}.
The corresponding energy density $\rho_c=3Q/4\pi R_H^3=3H^2c^2/8\pi G$ reaches closure density in the limit of a de Sitter universe.

In the face of a cosmological horizon ${\cal H}$ at $R_H$, therefore, the cosmological vacuum assumes thermodynamic properties by (\ref{EQN_Q}) very similar to those of black holes \citep{gib77}, satisfying $Q\equiv Mc^2$ by Clausius' integral for a Schwarzschild black hole of mass $M$ (e.g. \citep[][]{van24c}).
Accordingly, the vacuum of a Big Bang quantum cosmology is distinct from that of general relativity (GR), which assumes the asymptotic null-infinity ${\cal N}$ of Minkowski spacetime. 
{Given ${\cal N}$, the strong  field limit of GR notably predicts black holes including Kerr black holes 
\citep{ker23} and their evaporation by accompanying outgoing radiation conditions \citep{haw75,van24c}. 
Yet, the same ${\cal N}$ is inconsistent with ${\cal H}$ in (\ref{EQN_F}), prohibiting its direct application to weak gravitation on the scale of $a_{dS}$.}
 
{As a result, FLRW cosmologies (\ref{EQN_F}) have a non-classical vacuum (\ref{EQN_Q}), arising from 
an infinite cosmological redshift of ${\cal H}$ in (\ref{EQN_F}) rather than the zero redshift of ${\cal N}$ in Minkowski spacetime.}
We hereby anticipate a Hubble expansion $H(z)$ driven by a dark energy (\ref{EQN_Q}) distinct from the case of a constant dark energy, known as $\Lambda$CDM \citep{ade14,agh20,tri24}.

{\em An IR-consistent cosmological vacuum.} 
As a first step, we recall the bare cosmological constant $\Lambda_0=8\pi Gc^{-4}\rho_0$ of quantum field theory inferred from the Planck energy density $\rho_0=\hbar c/l_p^4$ \cite{zel67,wei89}. 
This $\Lambda_0\sim 1/\hbar$ is UV-divergent in $1/\hbar$ -
a {\em primitive} necessitating an 
IR-consistent coupling to spacetime satisfying aforementioned Bekenstein bound \citep{van24e}.
The result is expected to be dynamical by the swampland conjectures \citep{agr18}. 
{A similar primitive is encountered in coupling matter by position to spacetime \citep{van24e}.}

To this end, we consider (\ref{EQN_F}) in spherical coordinates with radial coordinate $r$ and the reciprocal $\alpha_p\sim \hbar$ of $A_p = 4 \pi r^2$,
\begin{eqnarray}
\alpha_p A_p = 1,
\label{EQN_IR}
\end{eqnarray}
over $0\le r\le R_H$. 
{For $r<R_H$, $\alpha_p$ provides an IR-consistent coupling of matter to spacetime \citep{van24e}. In a unified treatment, the visible Universe covered by $r=R_H$ leaves}
\begin{eqnarray}
    \Lambda = \alpha_p \Lambda_0 = 2H^2/c^2.
    \label{EQN_LA}
\end{eqnarray}
This outcome is less than limit $\rho_\Lambda=\rho_c$ in de Sitter space in keeping with aforementioned Bekenstein bound. For a three-flat universe (\ref{EQN_F}), this points to an additional distribution of dark matter. 
Excluding the de Sitter limit as a physical state of the Universe, it further points to a dynamical dark energy anticipated by the swampland conjectures.

By the above, therefore, the vacuum of a Big Bang cosmology (\ref{EQN_F}) is inequivalent to that of GR, and hence $\Lambda$CDM. By IR-consistent coupling to gravitation, we anticipate a dynamical dark energy (\ref{EQN_LA}). We formalize (\ref{EQN_LA}) by a path integral formulation gauged in global phase by ${\cal H}$ \citep{van20,van21}. 

This new approach on the Hubble expansion of a non-classical cosmological vacuum appears opportune in light of the $H_0$-tension between the Local Distance Ladder and standard $\Lambda$CDM \cite{agh20,div21,rie98,wei22,rie22}.

{\em Dark energy from first principles.} 
An exact expression for $\Lambda$ in (\ref{EQN_LA}) derives from a path integral formulation with gauged propagator $e^{i\left(\Phi-\Phi_0\right)}$ \citep{van20}, where $\Phi=S/\hbar$ for an action $S$ and $\Phi_0$ is a global phase reference. 
In the face of ${\cal H}$ rather than ${\cal N}$, $\Phi_0=\Phi_0\left[{\cal H}\right]$ represents a boundary term $S_0=\hbar\Phi_0\left[{\cal H}\right]$ in the total action.
In a Big Bang cosmology (\ref{EQN_F}), $\Phi_0$ is inherently dynamic, taking us away from de Sitter as anticipated above. To proceed, $S_0$ can be absorbed in $S$ by a Lagrangian density $2\Lambda$ as $\Lambda=\lambda R$ - a multiple of the Ricci scalar tensor as a function of the running Friedmann scale $a$.  
In what follows, we consider $\lambda=1/6$ inferred from thermodynamic arguments \citep{van15b}. 
In four dimensions, we recognize $J=R/6$ as the trace of the Schouten tensor known for conformal symmetries \cite{sch11}. That is,
\begin{eqnarray}
\Lambda=J,
\label{EQN_LB}
\end{eqnarray}
where $J\equiv (1-q)H^2/c^2$ in canonical quantities $(H,q)$ \cite{van15b}. This recovers 
(\ref{EQN_LA}) when $q=-1$ and $\Lambda=0$ in the radiation-dominated limit ($q=1$) prior to the BAO. 
By (\ref{EQN_LB}), $J$CDM is inequivalent to $\Lambda$CDM, which assumes $\Lambda$ to be frozen.

{\em Hubble expansion of $J$CDM.} 
With (\ref{EQN_LB}), we obtain an analytic solution
$H(z)=H_0h(z)$ \cite{van21},
\begin{eqnarray}
    h(z) = \frac{\sqrt{1+\frac{3}{2}\Omega_{K,0}Z_4(z)+\frac{6}{5}\Omega_{M,0}Z_5(z)+\Omega_{r,0}Z_6(z)}}{1+z},
    \label{EQN_HJ}
\end{eqnarray}
where $Z_n=\left(1+z\right)^n-1$ $(n\ge 1)$. 
Here, $\Omega_{M,0}$ and $\Omega_{r,0}$ denote the dimensionless densities of matter and, respectively, radiation at $z=0$, normalized to the present closure density $\rho_{c,0}$. $\Omega_{K,0}<0 \,(>0)$ is the density of positive (negative) curvature.
In the same notion, $\Lambda$CDM satisfies 
$H(z) = H_0E(z)$ with $E(z)=\sqrt{1+\Omega_{K,0}Z_2(z)+\Omega_{M,0}Z_3(z)+\Omega_{r,0}Z_4(z)}$.

At late times in $J$CDM, when radiation can be neglected, matter and dark energy densities satisfy $\Omega_M=\frac{1}{3}(q+2)$ and, respectively, $\Omega_\Lambda=\frac{1}{3}(1-q)$ and $q^\Lambda = \frac{3}{2}\Omega_M^\Lambda - 1$ \cite{van21}. 
The dark energy equation of state between pressure and energy, $p_\Lambda=w\rho_\Lambda$, satisfies $w=(2q-1)/(1-q)$ distinct from $w\equiv -1$ in $\Lambda$CDM \citep{van17b,col19}.
In the matter-dominated era ($q=1/2$, $w=0$), $\Omega_M=5/6$ and $\Omega_\Lambda=1/6$ conspire to preserve closure density at zero total pressure.
This reduces $\Omega_{M}$ partaking in large-scale structure formation in $J$CDM to 5/6 times the matter density in $\Lambda$CDM, i.e.:
\begin{eqnarray}
\Omega_{M,0}=\frac{5}{6}\Omega_{M,0}^\Lambda. 
\label{EQN_scaling1}
\end{eqnarray}

\begin{figure}[h]
\centerline{\includegraphics[scale=0.175]{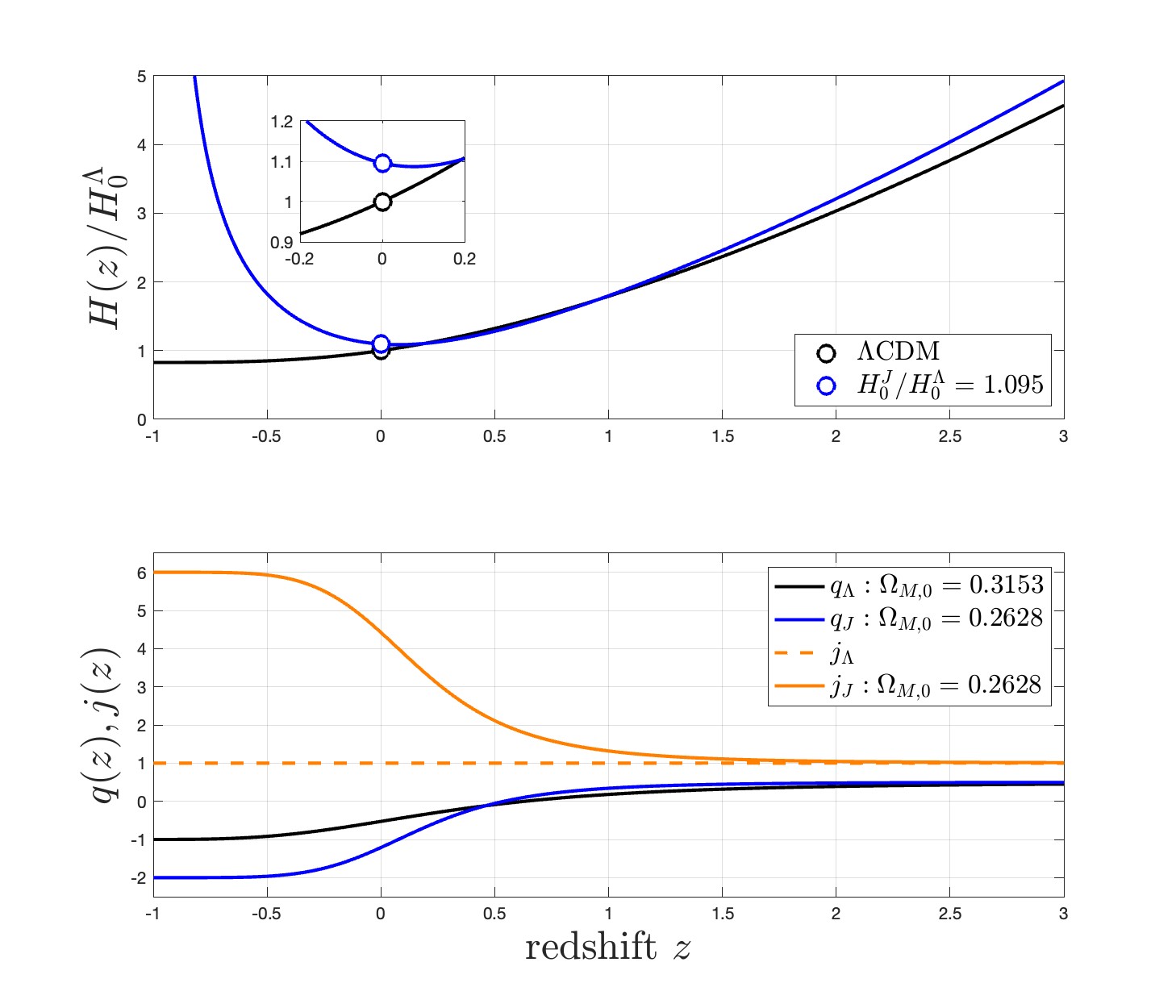}} 
\centerline{\includegraphics[scale=0.150]{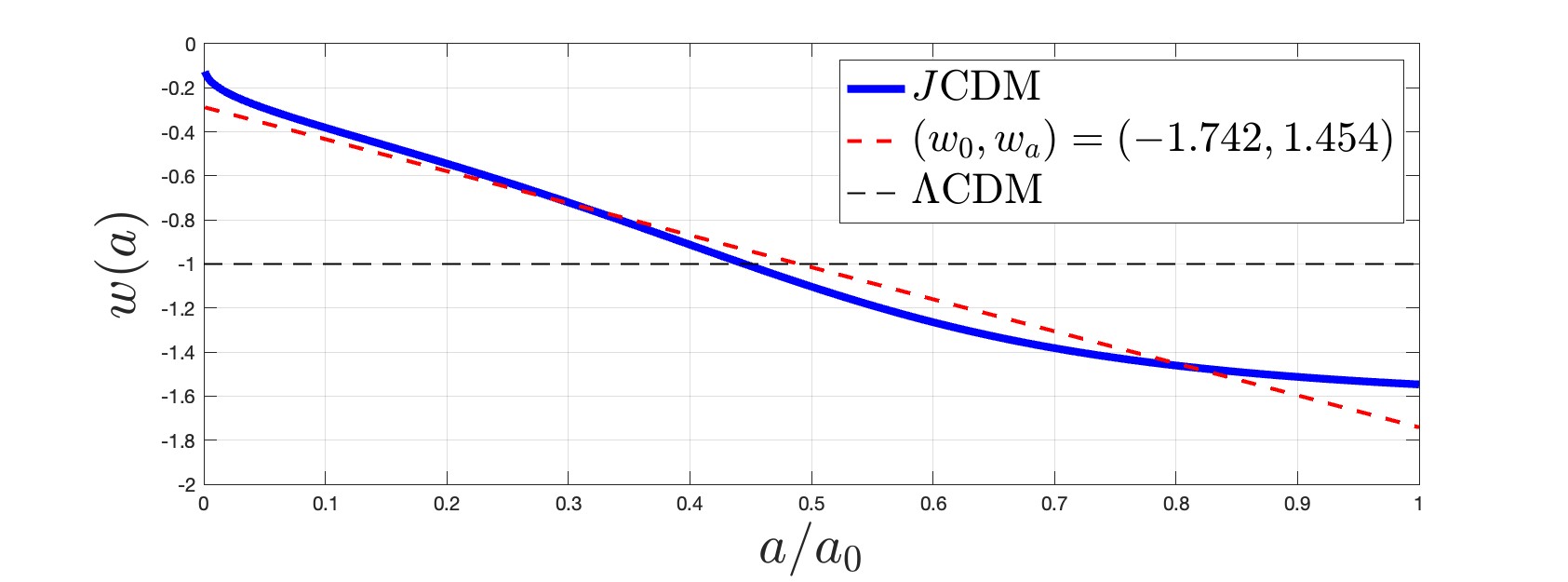}} 
    \vskip-0.003in
    \caption{
    (Top panel.) 
    $H(z)$ normalized to $H^\Lambda_0(z)$ of $\Lambda$CDM, highlighting sensitivity of $H_0$ in extrapolating low-$z$ data from the Local Distance Ladder data to $z=0$. This 
    extrapolation is sensitive to the shape of the graph, determined by the equation of state parameter $w$ of dark energy relating pressure and energy $p_\Lambda=w\rho_\Lambda$. Here, $w=(2q-1)/(1-q)$ for $J$CDM \citep{van17b,col19} and $w\equiv -1$ for $\Lambda$CDM.
    (Middle panel.) 
    $J$CDM departs from $\Lambda$CDM noticeably at late times.
    (Lower panel.) $H(z)$ in $J$CDM can be emulated by the Hubble expansion of $w(a)\Lambda$CDM with corresponding {nonlinear function} $w(a)$ (blue solid curve). 
    {For illustrative purposes, included} is the linear trend line (dashed red line)
    $w= w_0 + w_a(1-a)$.
    }
    \label{fig_JL}
\end{figure}

\begin{figure}[h]
\centerline{\includegraphics[scale=0.19]{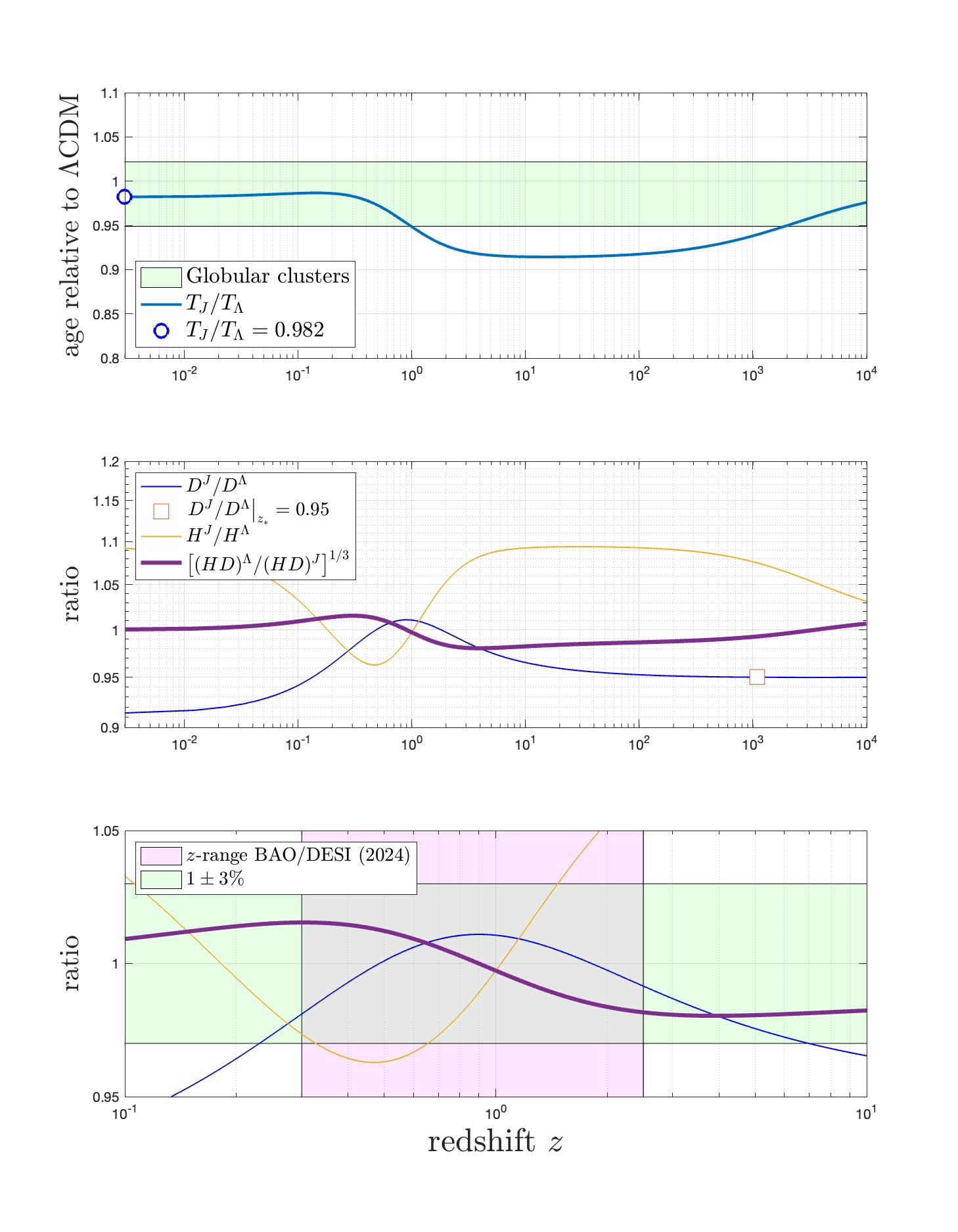}} 
    \vskip-0.0in
    \caption{
    (Upper panel.) The age $T_J$ of the Universe in $J$CDM relative to $T_\Lambda$ in $\Lambda$CDM, against the 
    astronomical age of the Universe inferred from the ages of the oldest stars in Globular Clusters \citep{val20}. In $J$CDM, the Universe is slightly younger than in $\Lambda$CDM by a mere 1.8\% (open circle).
    (Middle panel.) Shown are ratios of canonical quantities between $J$CDM and $\Lambda$CDM, namely $D_A^J/D_A^\Lambda=D_L^J/D_L^\Lambda\equiv D^J/D^\Lambda$ of angular distances $D_A=D_M/(1+z)$ and luminosity distances $D_L=(1+z)D_M$ in terms of the comoving distance $D_M$ (blue curve), of the Hubble expansion (yellow curve) 
    {and the cube root of $HD$ relevant to the recent DESI measurements of the BAO \citep{ada24} by (\ref{EQN_R}).
    (Lower panel.) Zoom-in of middle panel, highlighting consistency with {\em Planck}-$\Lambda$CDM within 2\% (green strip), consistent with a few percent uncertainty in DESI measurements.}
    }   
    \label{fig_TJ}
\end{figure}

Fig. \ref{fig_JL} shows the late-time Hubble expansion in $J$CDM alongside $\Lambda$CDM. By the distinct slope and curvature in the two graphs at the present epoch, we anticipate a tension in $H_0$ when extrapolating $H(z)$-data over $z>0$ from the Local Distance Ladder to $z=0$ \citep{abc21}.
This distinction is seen to be at late times $z\lesssim 1$ by $j\equiv \dddot{a}a^2/\dot{a}^3$
$\left( j(z)=q(z)(2q(z) + 1) + (1 + z)q^\prime(z)\right)$ with the property that $j\equiv 1$ for $\Lambda$CDM.
Due to this late-time transition, the age of the universe in $J$CDM remains close to that of $\Lambda$CDM shown in Fig. \ref{fig_TJ}, alongside the angular and luminosity distances relative to $\Lambda$CDM.

{\em Anchoring in the BAO.} In the radiation dominated epoch, $J$CDM and $\Lambda$CDM share the asymptotic expansion $H(z)\sim H_0 \sqrt{\Omega_{r,0}}(1+z)^{2}$. 
This suggests anchoring $J$CDM in early cosmology by the sound horizon in the surface of last scattering (SLS) at redshift $z_*\simeq1090$ according to the 
{\em Planck}-$\Lambda$CDM analysis of the CMB:
\begin{eqnarray}
    \theta_*=\theta_*^\Lambda,
    \label{EQN_theta}
\end{eqnarray}
where the superscript $J$ on the left hand-side is understood. 
Here, $\theta_*$ defined by the angle \cite{ade14,agh20,tri24}
\begin{eqnarray}
100\theta_* \equiv \frac{r_*}{D_*} = \left( 1.04092 \pm 0.00030\right) \,{\rm rad},
\label{EQN_BAO}
\end{eqnarray}
that determines the location of the main peak in the power spectrum of the CMB. It represents $\theta_*=r_*/D_*$ in terms of the radius $r_*=\int_{z_*}^\infty c_sdz/H(z)$ 
and the comoving distance $D_*=c\int_0^{z_*} dz/H(z)$ \citep{agh20,jed21,pit24}. Crucially, $c_s=1/\sqrt{3\left(1+R\right)}$ is the sound speed in the primordial baryon-poor fluid, determined by the baryon-to-photon density ratio $\rho_b/\rho_\gamma$, where $R=3\rho_b/4\rho_\gamma$.

\begin{figure}[h]
\centerline{\includegraphics[scale=0.1975]{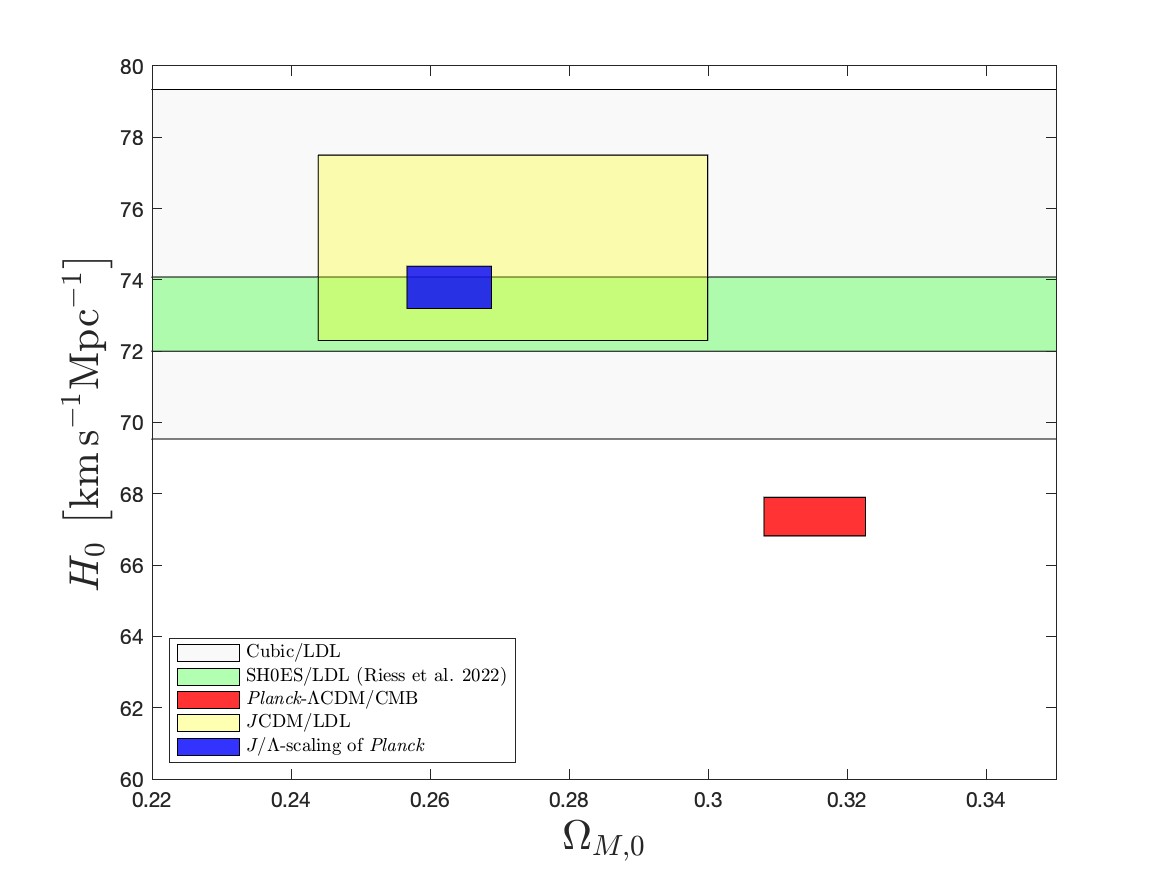}} 
\centerline{\includegraphics[scale=0.1975]{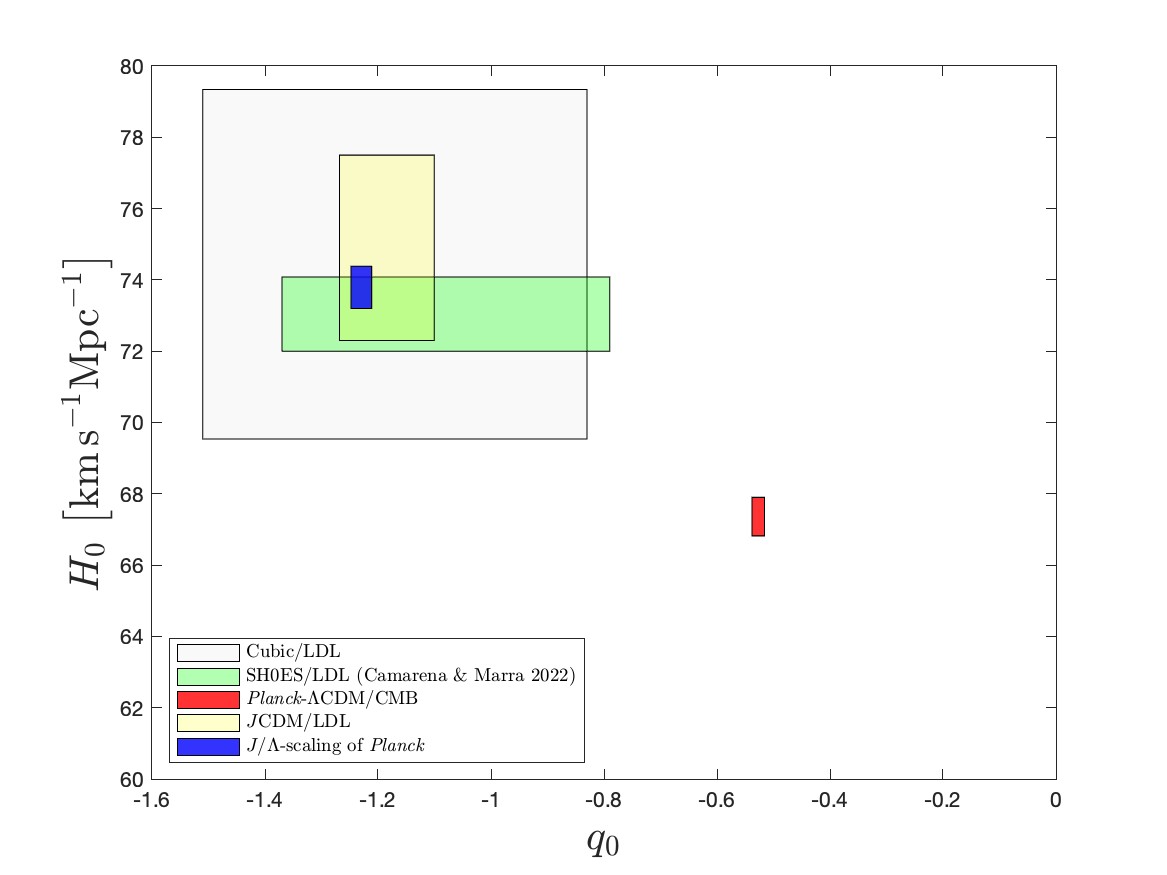}} 
    \vskip-0.0in
    \caption{
    (Top panel.) $\left(\Omega_{M,0},H_0\right)$-plane of Table 1, showing {\em Planck}-$\Lambda$CDM (red), SH0ES/LDL (green), $J$CDM fit to the LDL (yellow) and $J/\Lambda$-scaling of {\em Planck} (blue) following (\ref{EQN_scaling1}-\ref{EQN_scaling2}) against a model-agnostic background provided by a cubic polynomial fit to LDL (light gray). 
    (Bottom panel.) The same shown in the $\left(q_0,H_0\right)$-plane.
    }
    \label{figH0q0}    
\end{figure}

\begin{table*}
{Table 1.} $J$CDM parameter estimates versus {\em Planck}, the Local Distance Ladder (LDL) by the SH0ES collaboration \citep{rie22} and tabulated $H(z)$ data of \cite{far17}.
Results are expressed by
$\left(H_0,q_0\right)$, including $\Omega_{M,0}$ except in model-independent analyses of the LDL. 
Results by $J/\Lambda$-scaling are based on anchoring in the BAO (\ref{EQN_theta}), and, independently, from LDL. 
All results refer to three-flat cosmologies.
$S_8$ estimates derive from CAMB \cite{lew00}.
$H_0$ is in ${\rm km~s}^{-1}{\rm Mpc}^{-1}$ and 
$T_U$ is in Gyr.
\begin{eqnarray*}
\renewcommand\arraystretch{1.5}
\begin{array}{|c||cc|ccccc|ccc|}
    \hline
    && {\sc Planck}/{\rm CMB}^a & & J/\Lambda{\rm-scaling/BAO}^b & {} & J{\rm CDM}/{\rm LDL}^c & & {\rm\sc Cubic}/{\rm LDL}^d & & {\rm\sc SH0ES/LDL}^e  \\ 
    \hline \hline
    H_0 & & 67.36 \pm 0.54 & & 73.79\pm 0.59 & {} &   74.9\pm2.60 & &  74.44\pm4.9 & &  73.04 \pm1.04  \\
    q_0 & &-0.5273 \pm 0.011 & &-1.21\pm 0.014 &{}  &   -1.18\pm0.084 & & -1.17\pm0.34 & & -1.08 \pm0.29 \\
     \Omega_{M,0} & & 0.3153 \pm 0.0073 & & 0.2628\pm 0.0061 & {} & 0.2719\pm0.028 & &  - & &  - \\
    \hline
    S_8 && 0.832\pm 0.013 & &  0.756 \pm 0.012 && && && \\ 
    T_U  && 13.797\pm 0.023 & & 13.44 \pm 0.022 &&  && && \\
    \hline
\end{array}
\label{EQN_T1}
\end{eqnarray*}
$a$. {\em Planck}, Eq. (27) of \cite{agh20}.
$b$. Scaling relations (\ref{EQN_scaling1}-\ref{EQN_scaling2}). 
$c$. $J$CDM fit \citep{van17b} to tabulated $H(z)$ data of \citep{far17}.
$d$. Cubic polynomial fit \citep{van17b} to tabulated $H(z)$ data of \citep{far17}.
$e$. $H_0$ from \cite{rie22} and $q_0$ from \cite{cam20}. 
\end{table*}

To this end, we let (\ref{EQN_HJ}) preserve the {\em Planck} un-normalized dimensionless densities $\Omega_{M,0}h^2=0.1431$ and $\Omega_{r,0}h^2=2.4661\times10^{-5}$ (by the CMB temperature today) in $h=H_0/100$\,km\,s$^{-1}$Mpc$^{-1}$.
With (\ref{EQN_scaling1}), the first predicts the $J/\Lambda$-scaling
\begin{eqnarray}
H_0 = \sqrt{\frac{6}{5}}H_0^\Lambda 
\label{EQN_scaling2}
\end{eqnarray}
and $q_0=\left(5q_0^\Lambda -1\right)/3$. 
By (\ref{EQN_BAO}-\ref{EQN_scaling2}), (\ref{EQN_scaling1}) preserves the matter density $\rho_{M,0}$ of $\Lambda$CDM: $\Omega_{M,0}$ appears lower in the face of $\rho_c=(6/5)\rho_c^\Lambda$.
By (\ref{EQN_scaling2}), the radiation-to-matter density ratio satisfies $\eta=\Omega_{r,0}h^2/\Omega_{M,0}h^2= \left(1.7238\pm0.03\right)\times10^{-4}$.

Anchored by (\ref{EQN_theta}), $J$CDM preserves $\theta_*^\Lambda$ of three-flat $\Lambda$CDM at $N_{\rm eff}^\Lambda=3.046$. {In keeping with a direct confrontation between $J$CDM and $\Lambda$CDM, both are here considered with essentially zero curvature. In a CMB-only analysis, {the original} {\em Planck}-$\Lambda$CDM analysis of the CMB suggests a slight positive curvature, the value of which varies appreciably with choice of data combination. {While complex, this is now recognized to be largely a lensing anomaly, resolved to some extent in subsequent revised {\em Planck} analyses \citep{div20,tri24}.} 
For instance, $\Omega_k < -0.007$ in TT, TW, EE+lowE in \citep{agh20}. Numerical root finding of (\ref{EQN_theta}) shows the correlation $\Delta N_{\rm eff} \simeq 0.2566 + 0.1640 \times \left(100\Omega_K\right)$, according to which 
$0.1<\Delta N_{\rm eff}<0.26$ for 
$-1\le 100\Omega_K\le 0$ in $J$CDM. 
In particular, a fiducial value $\Omega_k=-0.007$ 
yields $\Delta N_{eff} = 0.1418$ for $J$CDM, i.e., $N_{eff} = 3.1878$. This $\Delta N_{eff}$ is well within the uncertainty of CMB-only {\em Planck} $\Lambda$CDM-parameter estimates \citep{ade14,agh20,tri24} and the corresponding $N_{eff}$ is consistent with the DESI estimate $N_{eff}=3.20\pm 0.19$ \citep{ada24}.
This sensitivity analysis on (\ref{EQN_theta}) indicates insignificant deviations from the concordance model of a flat cosmology with no need for extra relativistic degrees of freedom.}
   
{\em Confrontation with the Local Distance Ladder.}
By (\ref{EQN_scaling1}), (\ref{EQN_scaling2})  
and anchored by (\ref{EQN_theta}-\ref{EQN_BAO}), 
$J$CDM predicts $h\simeq 0.74$ with $\Omega_{M,0}\simeq 0.26$. $J$CDM hereby resolves the $H_0$-tension between the BAO and late-time cosmology in $\Lambda$CDM with negligible change in the age of the Universe (Fig. \ref{fig_TJ}).

\begin{figure*}[ht]
    \centerline{\includegraphics[scale=0.24]{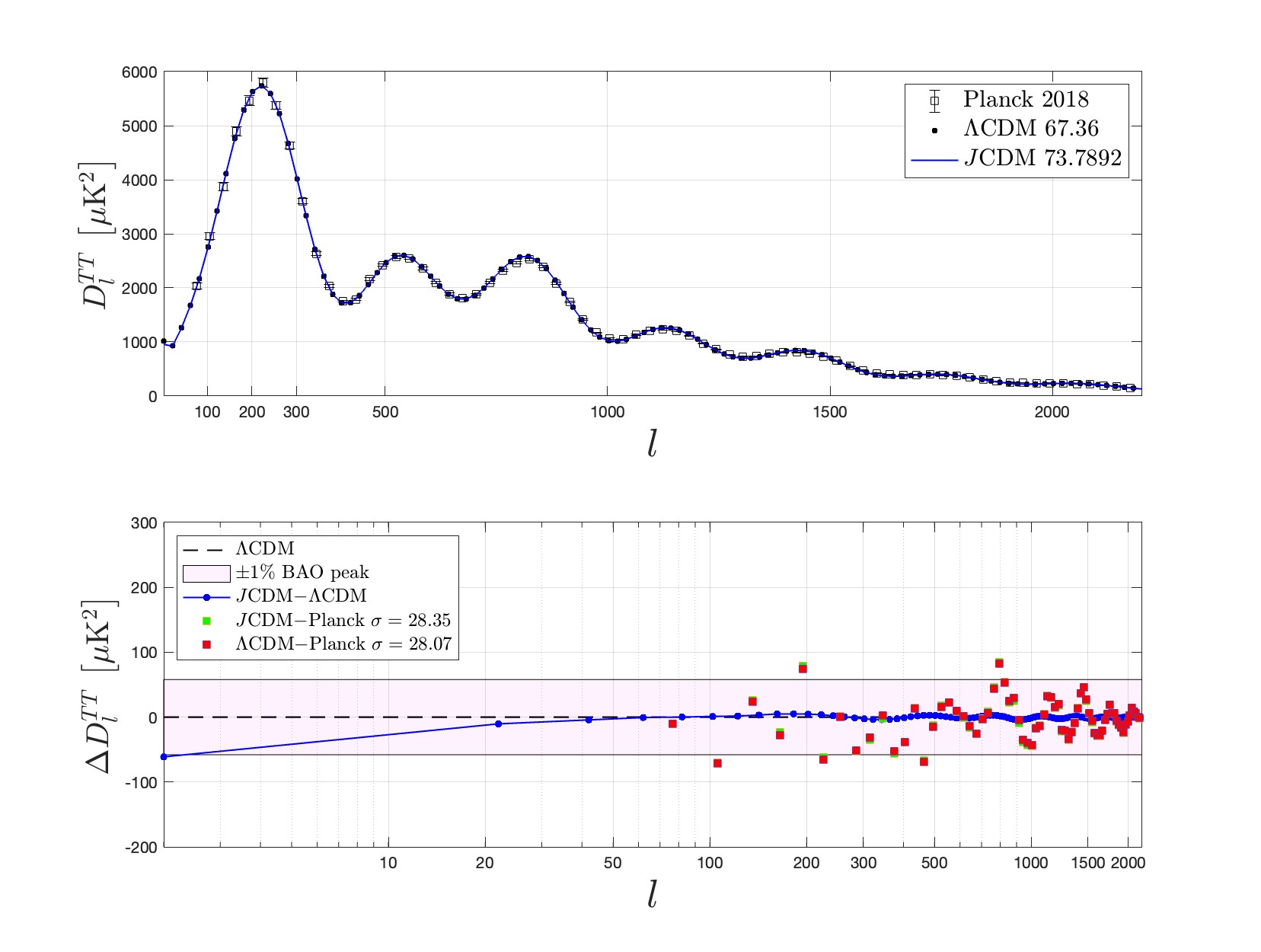}} 
    \vskip-0.0in
    \caption{
    $J$CDM and $\Lambda$CM model predictions by CAMB to binned {\em Planck} 2018 TT power spectrum of the CMB upon scaling $H_0$ by (\ref{EQN_scaling1}) and (\ref{EQN_scaling2}), preserving {\em Planck} $\Lambda$CDM values for all other parameters. 
    $J$CDM and $\Lambda$CDM power spectra are essentially the same except on larger scales ($l\lesssim 30)$, where $J$CDM falls below $\Lambda$CDM by about one percent.
    { The $\sigma$-values in the legends refer to STDs of $J$CDM and $\Lambda$CDM residuals to the binned {\em Planck} 2018 data.}
    }
    \label{figCAMB}
\end{figure*}

Similar but less pronounced departures from $\Lambda$CDM are seen in a minimal extension of CDM \citep{pit24}, predicting a low matter density $\Omega_{M,0}\simeq 0.267$ similar to ours with, however, a limited increase in the Hubble constant to $h=0.7187$. 

Table 1 and Fig. \ref{figH0q0} 
summarize the outcome of the confrontation of $J$CDM anchored in the BAO with late-time cosmology {probed by the Local Distance Ladder.}

{\em Planck CMB power spectrum.}
With $J/\Lambda$-scaling (\ref{EQN_scaling1}) and (\ref{EQN_scaling2}) and BAO anchoring (\ref{EQN_BAO}),
the CMB power spectrum in $J$CDM is expected to closely follow that of {\em Planck}.
Fig. \ref{figCAMB} shows a confrontation of $J$CDM and $\Lambda$CDM with binned $D_l^{TT}$ data of the {\em Planck} CMB power spectrum \citep{pla18} with small positive curvature
consistent with {\em Planck}-$\Lambda$CDM 
\citep{ade14,agh20,tri24}.
{Fig. \ref{figCAMB} is calculated by CAMB \citep{lew00} applied to $w(a)\Lambda$CDM emulating $J$CDM, using the nonlinear curve $w(a)$ (blue curve in Fig. \ref{fig_JL}) tabulated over 5000 data points covering the interval $0\le a \le 1$.}

For large scales, the $J$CDM power spectrum drops slightly below that of $\Lambda$CDM. This is a familiar feature of models beyond $\Lambda$CDM seen, e.g., in quintom cosmology \citep{cai10}, which is allowed given the relatively large cosmic variance at low $l$ \citep{dep10,ade14}.

On smaller scales, the matter clustering amplitude $\sigma_8$ on the scale of $8h^{-1}\,{\rm Mpc}$ is commonly expressed by the quantity $S_8=\sigma_8\sqrt{\Omega_{M,0}/0.3}$ \cite{agh20, div21,jed21}.
Preserving the {\em Planck}-$\Lambda$CDM parameter $\Omega_{M,0}h^2$, it satisfies
\begin{eqnarray} 
S_8\sim \sigma_8\,h^{-1}.
\label{EQN_S8}
\end{eqnarray}
While $\sigma_8$ slightly increases with $h$, $S_8$ hereby reduces by about 7\% in response to (\ref{EQN_scaling1}) (Table 1).

{\em Conclusions.} 
$J$CDM models the Hubble expansion of a non-classical vacuum of a Big Bang cosmology by first principles with no variation of fundamental constants \citep{van24f}. 
Its origin is an IR-consistent coupling of $\Lambda_0$ to spacetime (\ref{EQN_IR}-\ref{EQN_LB}) in keeping with the Bekenstein bound \citep{bek81} and formalized in $J$CDM based on a path integral formulation of (\ref{EQN_F}) with gauged global phase.

$J$CDM predicts $H_0$ to be larger than $H_0^\Lambda$ of $\Lambda$CDM by a factor of $\sqrt{6/5}$ and reduced matter density $\Omega_{M,0}$ (\ref{EQN_scaling1}) and (\ref{EQN_scaling2}). This $J/\Lambda$-scaling identifies the $H_0$-tension between the Local Distance Ladder and {\em Planck}-$\Lambda$CDM with relic heat in the vacuum, in a Hubble expansion anchored in the BAO of the {\em Planck} power spectrum of the CMB with reduced $S_8$ (Fig. \ref{figCAMB}, Table 1).

The dynamical dark energy (\ref{EQN_LB}) has a simple meaning in a finite content in heat: a relic of the Big Bang breaking time-translation invariance.
In turn, (\ref{EQN_HJ}) satisfies a new symmetry in the form of a T-duality in $a$ and $1/a$ \citep{van21}. 
This derives from the Hamiltonian energy constraint (the first Friedmann equation), now second-order in time by (\ref{EQN_LB}) in $J$CDM rather than first-order in time in $\Lambda$CDM, where $\Lambda$ is assumed to be frozen. 
This distinction highlights the observational significance of $q_0$ (Fig. \ref{fig_JL}).

{$J$CDM alleviates $H_0$-tension while anchored in the BAO (\ref{EQN_theta}-\ref{EQN_BAO}). 
A further test is provided by the BAO recently measured by DESI over a redshift range $0.3\lesssim z \lesssim 2.5$ \citep{ada24}, reported by the distance ratio $R\equiv D_V/(r_d z^{2/3})$ of the angle-averaged distance $D_V = (zD_M^2R_H)^{1/3}$ to the sound horizon at the baryon-drag epoch $r_d$ \citep{jed21,ada24}. 
Given the anchor (\ref{EQN_theta}), 
$J$CDM and {\em Planck}-$\Lambda$CDM results can be compared by the ratio
\begin{eqnarray}
\frac{R^J}{R^\Lambda} = \frac{\left(D_V/r_d\right)^J}{\left(D_V/r_d\right)^\Lambda}=\left(\frac{\beta^\Lambda}{\beta^J}\right)^{1/3},
\label{EQN_R}
\end{eqnarray}
where $\beta = HD_M/c$ denotes the ratio of transverse to line-of-sight comoving distance \citep{ada24}. 
Fig. \ref{fig_TJ} (lower panel, purple curve) shows (\ref{EQN_R}) to be within $1\pm 2\%$. $J$CDM hereby follows {\em Planck} to within 2\%, consistent with DESI given the uncertainty of $\sim 3\%$ in its BAO data \citep{ada24}.

{$J$CDM provides a novel framework for Hubble expansion described by the same six parameters as $\Lambda$CDM, and the results based on the constraint (\ref{EQN_theta}) and the $J/\Lambda$-scaling 
(\ref{EQN_scaling1}) and (\ref{EQN_scaling2}) of $\Omega_{M,0}$ and, respectively, $H_0$ suggest it may mitigate tensions between early and late-time cosmology.}

As shown in Fig. \ref{figH0q0}, a definitive discrimination between $J$CDM and $\Lambda$CDM may derive from an accurate measurement of $q_0$, to effectively determine the order of the first Friedmann equation for a Big Bang cosmology.

{Improved observational constraints on late-time cosmology} are expected from the BAO in present low-redshift galaxy surveys with the {\em Dark Energy Survey} \cite{ada24,des24} and the recently launched ESA mission {\em Euclid}. 
{These new probes} may confirm a crucial prediction: $q_0\simeq -1$ in $J$CDM (Fig. \ref{figH0q0}) distinct from the $\Lambda$CDM value $q_0\simeq -0.5$ \citep{cam20,van24d}. 
When preserving the fit to the {\em Planck} power spectrum of the CMB on par with $\Lambda$CDM (Fig. \ref{figCAMB}), a tension in $\left(H_0,q_0\right)$ would represent a definite signature of a non-classical vacuum beyond $\Lambda$CDM - beyond the Hamiltonian energy constraint of general relativity.
New constraints on the Hubble expansion are further expected from increasingly high-resolution maps of galaxy formation at cosmic dawn by the JWST \cite{eis23} and galaxy rotation curves tracing background cosmology \citep{van24d,van24e}.

{\bf Acknowledgments.} We gratefully acknowledge the anonymous reviewer for constructive comments, M.A. Abchouyeh for detailed discussions and the organizers of {\em Tensions in Cosmology}, Corfu 2023, for a stimulating meeting on the dark sector of the Universe.
This research is supported, in part, by 
NRF grant No. RS-2024-00334550 and 
the MSIT and ICT of Korea under the ITRC support program IITP-2024-00437191.


\begin{thebibliography}{99}

\bibitem[Abbot et al.(2024)]{des24} 
DES Collaboration: Abbott, T.M.C., Adamov, M., Aguena, M., 2024, arXiv:2402.10696 

\bibitem[Abchouyeh \& van Putten(2021)]{abc21}
Abchouyeh, M.A., \& van Putten, M.H.P.M., 2021 
{\em Phys. Rev. D}, {\bf 104}, 083511 

\bibitem[Adame et al.(2024)]{ada24}
DESI Collaboration: Adame, A. G., Aguilar, J., Ahlen, S., et al., 2024, JCAP, to appear; arXiv:2404.03002  

\bibitem[Ade et al.(2014)]{ade14}
Ade, P.A.R., Aghanim, N., Armitage-Caplan, C., Arnoud, M., et al., 2014, {\em A\&A} {\bf 571}, A16 

\bibitem[Aghanim et al.(2020)]{agh20}
Aghanim, N., Akrami, Y., Ashdown, M., et al., 2020, {\em A\&A} {\bf 641}, A6 

\bibitem[Agrawal et al.(2018)]{agr18} 
Agrawal, P., Obied, G., Steinhardt, P.J., Vafa, C., 2018, {\em Phys. Lett. B} {\bf 784}, 271; 
Obied, G., Horiso, O., Spodyneiko, L., Vafa, C., 2018, arXiv:1806.08362v3; 
Garg, S.K., \& Krishnan, C., 2019, {\em JHEP} {\bf 11}, 075


\bibitem[Bekenstein(1981)]{bek81} 
Bekenstein, J.D., 1981, {\em Phys. Rev. D} {\bf 23}, 287

\bibitem[Cai et al.(2010)]{cai10}
Cai, Y.-F., Saridakis, E.N., Setare, M.R., \& Xia, J.-Q., 2010, {\em Phys. Rep.} {\bf 493}, 1 

\bibitem[Camarena \& Marra(2020)]{cam20} Camarena, D., \& Marra, V., {\em Phys. Rev. Research} {\bf 2}, 013028 (2020)

\bibitem[Col\'ain et al.(2019)]{col19}
Colg\'ain, E.\'O, van Putten, M.H.P.M., \&  Yavartanoo, H., 2019, {\em Phys. Lett. B} {\bf 793}, 126

\bibitem[de Putter et al.(2010)]{dep10}
Roland de Putter, R., Huterer, D., \&  Linder, E.V., 
{\em Phys. Rev. D}, {\bf 81}, 103513 (2010)

\bibitem[Di Valentino et al.(2020)]{div20}
Di Valentino, E., Melchiorri, A., \& Silk, J., 2020, {\em Nat. Astron.}, {\bf 4}, 196

\bibitem[Di Valentino et al.(2022)]{div21}
Di Valentino, E., Mena, O., Pan, S., Visinelli, L., Yang, W., Melchiorri, A., Mota, D.F., Riess, A.G., \& Silk, J., 2021, {\em Class. Quant. Grav.} {\bf 38}, 153001; 
Perivolaropoulos, L., \& Skara, F., 2022, {\em New Astron. Rev.} {\bf 95}, 101659

\bibitem[Eisenstein et al.(2023)]{eis23} 
Eisenstein, D.J., Willott, C., Alberts, S., et al., 2023, arXiv:2306.02465; Austin, D., Adams, N., Conselice, C.J., et al., 2023, {\em  ApJ}, {\bf 952}, L7 

\bibitem[Farooq et al.(2017)]{far17} Farooq, O., Madiyar, F.R., Crandall, S. and Ratra, B., 2017, {\em ApJ} {\bf 835}, 26

\bibitem[Gibbons \& Hawking(1977)]{gib77}
Gibbons, G.W., \& Hawking, S.W., 1977, {\em Phys. Rev. D} {\bf 15}, 2738

\bibitem[Hawking(1975)]{haw75}
Hawking, S.W., 1975, {\em Commun. Math. Phys.} {\bf 43}, 199

\bibitem[Jedamzik et al.(2021)]{jed21}
Jedamzik, K., Pogosian, L., \& Zhao, G.-B., 2021, {\em Comm. Phys.} {\bf 4}, 123

\bibitem[Kerr(2023)]{ker23}
Kerr, R.P., 2023, arXiv:2312.00841v1; 
$ibid$. 1963, {\em Phys. Rev. Lett.} {\bf 11}, 237; 
Kerr, R.P. \&  Schild, A., 1965, {\em Atti del Convegno sulla Relativita Generale: Problemi dell’Energia e Onde Gravitazionali}, 1–12, G. Barbèra Editore, Firenze;
Kerr, R.P., \& Schild, A., 1965, Proc. Symp. Appl. Math, R. Finn, Ed., {\em AMS} p.173;
Wiltshire, D.L., Visser, M., and Scott, S.M., 2009, {\em The Kerr Spacetime} (Camb. Univ. Press), p.38

\bibitem[Lewis et al.(2000)]{lew00}
Lewis, A., Challinor, A., \& Lasenby, A., 2000, {\em ApJ}, {\bf 473} 

\bibitem[Padmanabhan(2003)]{pad05}
Padmanabhan, T., 2003, {\em Phys. Rep.} {\bf 406}, 49 

\bibitem[Pitrou \& Uzan(2024)]{pit24}
Pitrou, C., \& Uzan, J.-P., 2024, {\em Phys. Rev. Lett.}, {\bf 132}, 191001

\bibitem[{\em Planck} PDR-3 (2018)]{pla18}
{\em Planck} Public Data Release 3, 2018, 
https://irsa.ipac.caltech.edu/data/Planck/release${_-}$3/ ancillary-data/

\bibitem[Riess et al.(1998)]{rie98}
Riess, A.G., Filippenko, A.V., Challis, P., et al., 1998, {\em ApJ} {\bf 116}, 1009; Perlmutter, S., Aldering, G., Goldhaber, G., et al., 1999, {\em ApJ} {\bf 517}, 565;
Riess, A.G., Macri, L.M., Hoffmann, S.L., et al., 2016, {\em ApJ} {\bf 826} 56; Riess, A.G., 2020, {\em Nat. Rev. Phys.} {\bf 2}, 10; Verde, L., True, T., Riess, A.G., 2019, {\em Nat. Astron.} {\bf 3}, 891;
Khetan, N., Izzo, L., Branchesi., M., et al., 2021, {\em A\&A} {\bf 647}, A72 

\bibitem[Riess et al.(2022)]{rie22} 
Riess, A.G., Yuan, W., Macri, L.M., et al., 2022, {\em ApJ} {\bf 934}, L7

\bibitem[Sandage(1961)]{san61} 
A. Sandage, 1961, {\em ApJ} {\bf 133}, 355 

\bibitem[Schouten(2011)]{sch11} 
Schouten, J.A., 2011, {\em Tensor analysis for physicists} (Dover)); 
Wald, R.M., 1984, {\em General Relativity} (Univ Chicago Press); 
Tugov, I.I., 1969, {\em Ann. Inst. Henri Poincar\'e A}, XI, 207

\bibitem[Tristan et al.(2024)]{tri24}
Tristam, M., Banday, A.J., Douspis, M., et al., 2024, {\em A\&A} {\bf 682} 

\bibitem[Valcin et al.(2020)]{val20} 
Valcin, D., Bernal, J.L., Jimenez, R., Verde, L., Wandelt, B.D., 2020, {\em JCAP} {\bf 12}, 002

\bibitem[van Putten(2015)]{van15b}
van Putten, M.H.P.M., 2015, {\em MNRAS} {\bf 450}, L48 

\bibitem[van Putten(2017b)]{van17b}
van Putten, M.H.P.M., 2017, {\em ApJ}, {\bf 848}, 28 

\bibitem[van Putten(2020)]{van20} 
van Putten, M.H.P.M., 2020, {\em MNRAS} {\bf 491}, L6 
\bibitem[van Putten(2021)]{van21}
van Putten, M.H.P.M., 2021, {\em Phys. Lett. B} {\bf 823}, 136737
\bibitem[van Putten(2024a]{van24c}
van Putten, M.H.P.M., 2024a, {\em Class. Quant. Grav.}, {\bf 41}, 06LT01
\bibitem[van Putten (2024b)]{van24d} 
van Putten, M.H.P.M., 2024b, {\em Phys. Dark Univ.}, {\bf 43},  101417 
\bibitem[van Putten(2024c)]{van24e}
van Putten, M.H.P.M., 2024c, {\em ChJPh} {\bf 91}, 377 (2024); Proc. {\em Tensions in Cosmology}, Corfu2024, {\em PoS} {\bf 463} https://doi.org/10.22323/1.463.0208 
\bibitem[van Putten(2024f)]{van24f}
van Putten, M.H.P.M., 2024, {\em Results in Physics}, {\bf 57}, 107425

\bibitem[Weinberg \& White(2022)]{wei22} 
Weinberg, D.H. \& White, M., {\em in} Rev. Particle Physics, R.L. Workman et al., 2022, {\em PTEP} {\bf 083C01} 

\bibitem[Weinberg(1998)]{wei89}
Weinberg, S., 1989, {\em Rev. Mod. Phys.} {\bf 61}, 1 (1989)


\bibitem[Zeldovich(1977)]{zel67} 
Zeldovich, Ya.-B., 1967, {\em JETP Lett.} {\bf 6}, 316

\end{thebibliography}
\end{document}